\documentclass[useAMS,usenatbib]{mnras}
\usepackage[T1]{fontenc}
\usepackage{aecompl}

\def\msun{\hbox{M$_{\odot}$}~}

\def\teff{T$_{eff}$}

\def\logg{$\log (g)$}

\def\vt{$v_t$}
\def\iron{[Fe/H]}
\usepackage{color}
\usepackage{url}
\usepackage{enumitem}
\usepackage{placeins}
\usepackage{graphicx}
 \usepackage{times}
  \usepackage{rotating}
  \usepackage{txfonts}
\defcitealias{Pancino2017}{P17}

\title{Chemical Study of the Metal-rich Globular Cluster NGC 5927}
\author[A. Mura-Guzm\'an. et al.]{
  A. Mura-Guzm\'an.$^{1}$\thanks{E-mail: aldomura@udec.cl},
  S. Villanova$^1$,
  C. Mu\~noz$^{1,2}$,
  B. Tang$^{3,1}$
\\
  $^1$Universidad de Concepci\'on, Departamento de Astronom\'{\i}a, Casilla 160-C, Concepci\'on, Chile\\
  $^2$ European Southern Observatory, Vitacura, Santiago, Casilla 19001, Chile\\
  $^3$ School of Physics and Astronomy, Sun Yat-sen University, Zhuhai 519082, China
  }
\date{}

\pagerange{\pageref{firstpage}--\pageref{lastpage}} \pubyear{2017}

\def\LaTeX{L\kern-.36em\raise.3ex\hbox{a}\kern-.15em
    T\kern-.1667em\lower.7ex\hbox{E}\kern-.125emX}

\begin{document}

\label{firstpage}

\maketitle 

\begin{abstract} 
Globular Clusters (GCs) are natural laboratories where stellar and chemical evolution can be studied in detail. In addition, their chemical patterns and kinematics can tell us wich Galactic structure (Disk, Bulge, Halo or extragalactic) the cluster belongs to. NGC 5927 is one of most metal-rich GCs in the Galaxy and its kinematics links it to the Thick Disk.
We present abundance analysis based on high resolution spectra of 7 giant stars. The data were obtained using FLAMES/UVES spectrograph mounted on UT2 telescope of the European Southern Observatory.
 The principal motivation of this work is to perform a wide and detailed chemical abundance analysis of the cluster and look for possible Multiple Populations (MPs).
We determined stellar parameters and measured 22 elements corresponding to light (Na, Al), alpha (O, Mg, Si, Ca, Ti), iron-peak (Sc, V, Cr, Mn, Fe, Co, Ni, Cu, Zn) and heavy elements (Y, Zr, Ba, Ce, Nd, Eu). We found a mean iron content of [Fe/H]=-0.47 $\pm$0.02 (error on the mean). We confirm the existence of MPs in this GC with an O-Na anti-correlation, and moderate spread in Al abundances. We estimate a mean [$\alpha$/Fe]=0.25 $\pm$0.08. Iron-peak elements shows no significant spread. The [Ba/Eu] ratios indicate a predominant contribution from SNeII for the formation of the cluster.
\end{abstract}


\section{INTRODUCTION}

Galactic Globular Clusters (GGCs) are considered as natural laboratories for the study of different stellar evolution processes, since they are among the oldest objects in the universe. We can compare GCs with the different components of the Milky Way looking for inter-relationships in their chemistry and/or kinematics. The Galaxy is made-up by at least 3 main structures: The Halo, the Bulge and the Disk. This last one is suggested to have there major components: Thick Disk and Thin Disk \citep{Gilmore1983}. The differences between the Galaxy components relies on their metallicity, abundance enhancements, age and kinematics among others.\\

When we talk about the Galactic Bulge, we are referring to an important structure of the Milky Way which mass estimations goes from the 10\% \citep{Oort1977} to the 25\% \citet{Sofue2009} of the total stellar mass in the Milky Way. Comparing the Bulge field stars with metal-rich GCs, \citet{Ortolani1995} and \citet{Zoccali2003} have estimated an age about $t=10 \pm 2.5$ Gyr for the bulk of the metal-rich population of the Galactic Bulge and its kinematics behaviour is dominated by a rotationally supported system  with a rotational peak of $\sim$75 km s$^{-1}$ \citep{Minniti1996,Beaulieu2000,Rich2007}, in addition to a high velocity dispersion \citep{Minniti1996} which decreases along the distance from the Galactic center. 
 Meanwhile, the Thin and Thick Disk have different stellar populations where \citet{Gilmore1983} observed for the first time these two populations by studying the stellar density at different Galactic latitudes towards the Galactic North pole. Many observational evidence \citep{Gratton2000,Mashonkina2001,Reddy2003,Reddy2006,Adibekyan2012} indicates that both sub-structures have different chemical origins and have passed through different chemical evolution. The Thick disk host old stars within 8 to 10 Gyr while the Thin disk stars show younger ages \citep{Bensby2003,Schuster2006}.\\

Not long ago, stars in GCs were originally considered coeval \citep{Benzini1986} and initially chemically homogeneous. Today, we know that GCs host Multiple Populations (MPs) in their stellar content and both photometric and spectroscopic methods are used in order to understand this phenomenon. Using spectroscopy we can observe spreads in the light element contents (C, N, O, Na, Mg, Al) forming patterns between them as correlations or anti-correlations. These signatures come from different reactions \citep{Denisenkov1989,Langer1993} which are simultaneously active, depleting some light elements and enhancing others. These reactions are carried out by proton capture during Hydrogen burning at high temperature ($\sim 15 \times 10^6$K for CNO chain, $\sim 30 \times 10^6$K for NeNa chain and $\sim 70 \times 10^6$K for MgAl chain) achieved in relatively massive stars (M $\geqslant$ 2\msun). The feature best studied is the Na-O anti-correlation \citep{Carretta2009a,Carretta2009b}.	Every GGC we know so far follows this pattern with the only exception of Ruprecht 106, but the origin of this cluster may be extragalactic \citep{Villanova2013}. This anti-correlation between Na and O is probably due to the evolution of massive stars in early stages of the GC. A First Generation (FG) of stars (i.e. Na-poorer O-rich) polluted the intra-cluster gas left behind with processed material and then a second generation (SG) of stars was formed being Na-richer and O-pooer \citep{Caloi2011}. There are several candidates for this light element polluter: intermediate mass AGB stars \citep{Dantona2002}, fast rotating massive MS stars \citep{Decressin2007} and massive binaries \citep{Mink2009}.
On the other hand, heavy elements are produced by neutron capture. This process can be slow or rapid and is carried out in AGB stars or SNeII explosions respectively. The relative abundances of neutron capture elements give us an idea about the process that contributed more in the chemical evolution of the cluster as we will discuss later.

In this paper we present a chemical study of the GC NGC 5927 based on high resolution spectra taken from UVES. This cluster is very old with an age of 12.25 Gyr \citep{Dotter2010}. This means that it was formed during the earliest stages of the Galaxy formation and could give us a hint about how the initial material got processed chemically. About kinematics, \citet{Casetti-Dinescu2007} and \citet{Allen2008} concluded that the orbital parameters of NGC 5927 are consistent with a rotationally supported system, suggesting the GC belongs to the Thick Disk. In terms of chemical abundances instead, there is a lack of studies using high resolution spectra. Many attempts have been made in order to estimate its metallicity based on Ca$_{II}$ IR triplet measurements \citep{Rutledge1997}. Then, in the catalogue performed by \citet[ed. 2010]{Harris1996}, a value of [Fe/H]=-0.49 dex was obtained by averaging metalicities derived from \citet{Armandroff1988}, \citet{Francois1991} and \citet{Carretta2009}. Finally, the most recent study about chemical analysis related to NGC 5927 was performed by \citet[Here and after P17]{Pancino2017} where their study was focused in the Mg-Al anti-correlation in 9 GCs including NGC 5927 using the GAIA-ESO Survey (GES). \citetalias{Pancino2017} used 9 stars for the abundance analysis in NGC 5927 which 7 of them correspond to the same stars used in this work. Nevertheless, the data collected in both works correspond to different observations and data reduction. The elements presented in \citetalias{Pancino2017} are Fe, O, Na, Mg and Al which we are going to compare with our measurements.\\

In Section 2 we describe the observations and data reduction performed for the study. Section 3 is a description of the methods used in order to derive the chemical abundances and the errors analysis. In Section 4 our results are presented and compared with different components of the Galaxy and other GCs and finally in Section 5 we give a summary, present our findings and proffer our conclusions.

\section{OBSERVATIONS AND DATA REDUCTION}

\begin{figure}
  \centering
    \includegraphics[width=.95\linewidth]{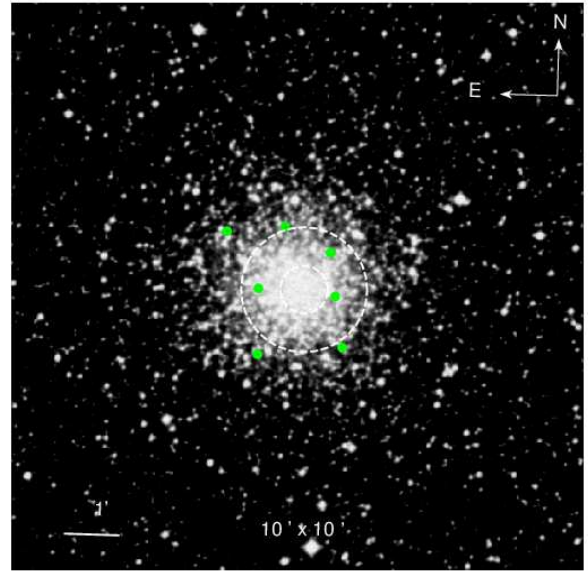}
    \caption{A $10\arcmin \times 10\arcmin$ digitized sky survey image centered on NGC 5927. North is up and east to the left. The green symbols show the location of the spatial distribution of the 7 stars analysed. The dashed circles correspond to the core radius, r$_c=0.42\arcmin$, and half mass-radius, r$_h=1.10\arcmin$ (\citealt{Harris1996}, 2010 ed).}
     \label{Fig:5927stars}
\end{figure}

\begin{figure}
  \centering
    \includegraphics[width=.95\linewidth]{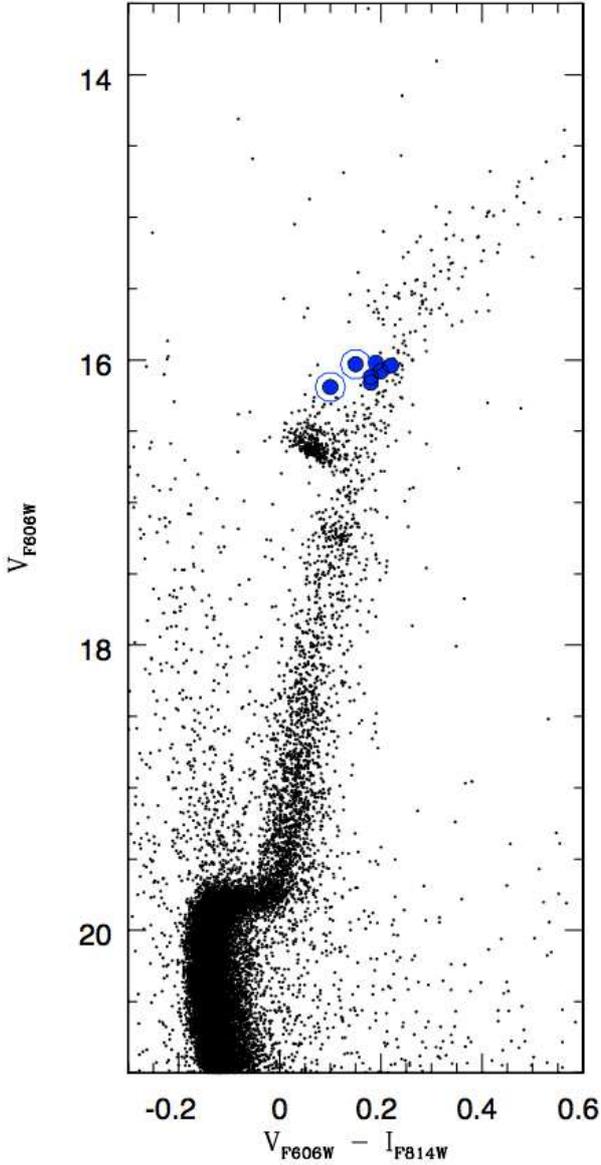}
    \caption{CMD of NGC 5927 adapted from \citet{Simmerer2013}. Blue filled circles represents the 7 stars under study.}
     \label{Fig:cmd}
\end{figure}

The dataset consist of 7 high resolution spectra (R $\sim 40000$) of 7 giant stars from NGC 5927 using the Ultraviolet Echelle Spectrograph (UVES) spectrograph \citep{UVES}. The central wavelength was centered in 580 nm and the spectral range covers between 480 nm and 620 nm (CCD$\#$3, filter SHP700, red arm only). The data was obtained from the ESO Archive\footnote{Observations were made in ESO facilities on VLT-UT2 telescope, Cerro Paranal, under program ID 079.B-0721(A).} as public data. The selection of targets was made using a CMD based on observations with the Advanced Camera for Surveys (ACS) on-board Hubble Space Telescope (See figure \ref{Fig:cmd}). See \citet[Sections 3 and 4]{Simmerer2013} for more detailed explanation about the target selection. The spatial distribution of the observed stars is shown in figure \ref{Fig:5927stars} including the core radius and the half mass-radius. Also, \citet{Allen2008} calculated a tidal radius for NGC 5927 (not show in the figure \ref{Fig:5927stars}) using \citet{King1962} formula (R$_K$) and a reduced King's formula (R$_*$) in two different potential. In an axisymmetric potential R$_K$=52.1pc and R$_*$=61.5pc but under a barred potential R$_K$=49.8pc and R$_*$=57.6pc. From Figure \ref{Fig:cmd}, there seems to be two AGB stars (star $\#3$ and star $\#4$) in our data which we marked with a circle in plots. In the case of star $\#$3 shows atypical abundances values as discuss in the following sections. This star is the hottest one (See Table \ref{tab:stellar_params}), but received the same treatment to obtain the stellar parameters.

The signal-to-noise ratio (S/N) in our spectra is between 20-50 at $\sim 600$ nm. Table \ref{tab:SN} shows the S/N of each star in two different wavelength ranges. About data reduction, bias subtraction, flat-field correction and wavelength calibration were performed using the UVES pipeline version 5.7.0\footnote{Pipelines list on \url{http://www.eso.org/sci/software/pipelines/}}. The sky subtraction, and spectral rectification were performance using \textit{Sarith} and \textit{continuum} tasks in IRAF.

\begin{table} 
\centering
\begin{large}
\begin{tabular}{ccc}
\hline
\hline 
 Star & 604.7-605.3 [nm] & 606.8-607.6 [nm] \\
\hline
1 & 48.13 & 50.24\\
2 & 41.36 & 35.05\\
3 & 39.72 & 37.54\\
4 & 30.58 & 36.00\\
5 & 34.82 & 40.56\\
6 & 36.95 & 40.30\\
7 & 47.64 & 49.49\\
\hline
\hline
\end{tabular}
\end{large}
\caption{Estimation of the signal-to-noise ratio in each star between 604.7-605.3 nm and 606.8-607.6 nm.}
\label{tab:SN}
\end{table}

Radial velocities were measured using the \textit{fxcor} task in IRAF and a synthetic spectrum as a template. Radial velocities indicate that the 7 stars are members of the cluster and we found a mean radial velocity of $v_r=-102.45 \pm 3.50$ km s$^{-1}$. This value is in good agreement with \citet{Simmerer2013}, who found a mean radial velocity of $v_r= -104.03 \pm 5.03$ km s$^{-1}$ for the cluster using the same UVES data, but also GIRAFFE data (72 member stars).

\begin{table*}
\caption{Basic parameters of our star sample in NGC 5927.}
\begin{tabular}{lccccccccc}
\hline
\hline
ID& RA$_{fibre}$& DEC$_{fibre}$& V$_{F606W}$& I$_{F814W}$& $RV_H$& $T_{eff}$& $\log (g)$& $v_t$& [Fe/H]\\
\hline         
1& 232.02279& -50.67275& 16.04& 15.82& -105.279& 4519& 2.61& 1.10& -0.54\\
2& 231.99008& -50.66175& 16.16& 15.98& -101.163& 4500& 2.60& 1.21& -0.49\\
3$^{\dagger}$& 232.02242& -50.69211& 16.19& 16.09& -102.328& 4835& 2.86& 1.59& -0.44\\
4$^{\dagger}$& 232.01121& -50.65447& 16.03& 15.88& -105.656& 4422& 2.56& 1.40& -0.47\\
5& 231.98354& -50.68953& 16.02& 15.83& -102.252& 4430& 2.52& 1.22& -0.51\\
6& 232.03792& -50.65622& 16.08& 15.88& -105.538& 4376& 2.28& 1.06& -0.37\\
7& 231.98742& -50.67450& 16.12& 15.94& -94.948&  4358& 2.24& 0.96& -0.49\\
\hline
\end{tabular}
\begin{tabular}{lccccccccc}
$^{\dagger}$ AGB stars. & & & & & & & & & \\
\end{tabular}
\label{tab:stellar_params}
\end{table*}

\begin{figure}
  \centering
    \includegraphics[width=0.95\linewidth]{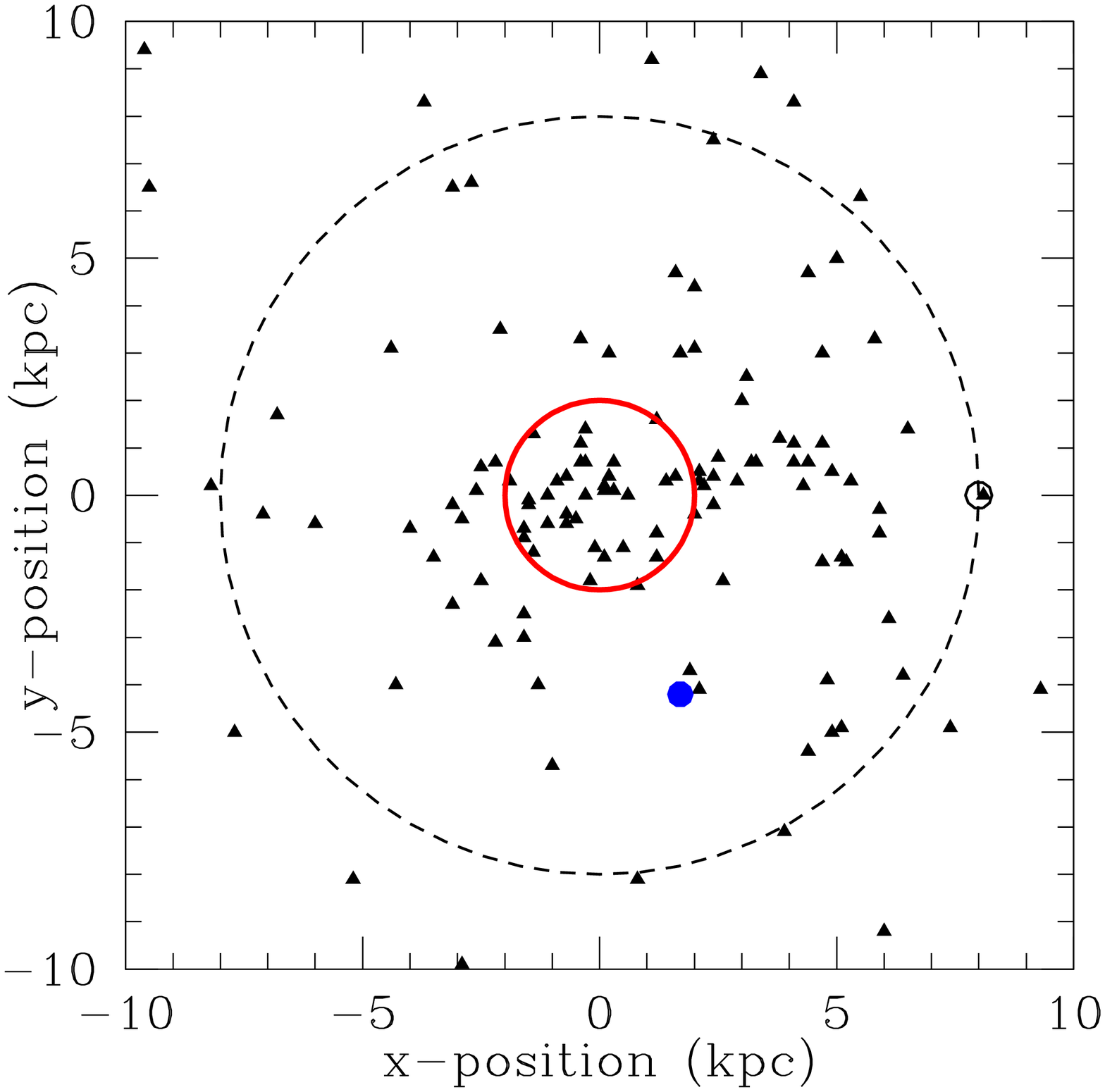}
    \includegraphics[width=0.95\linewidth]{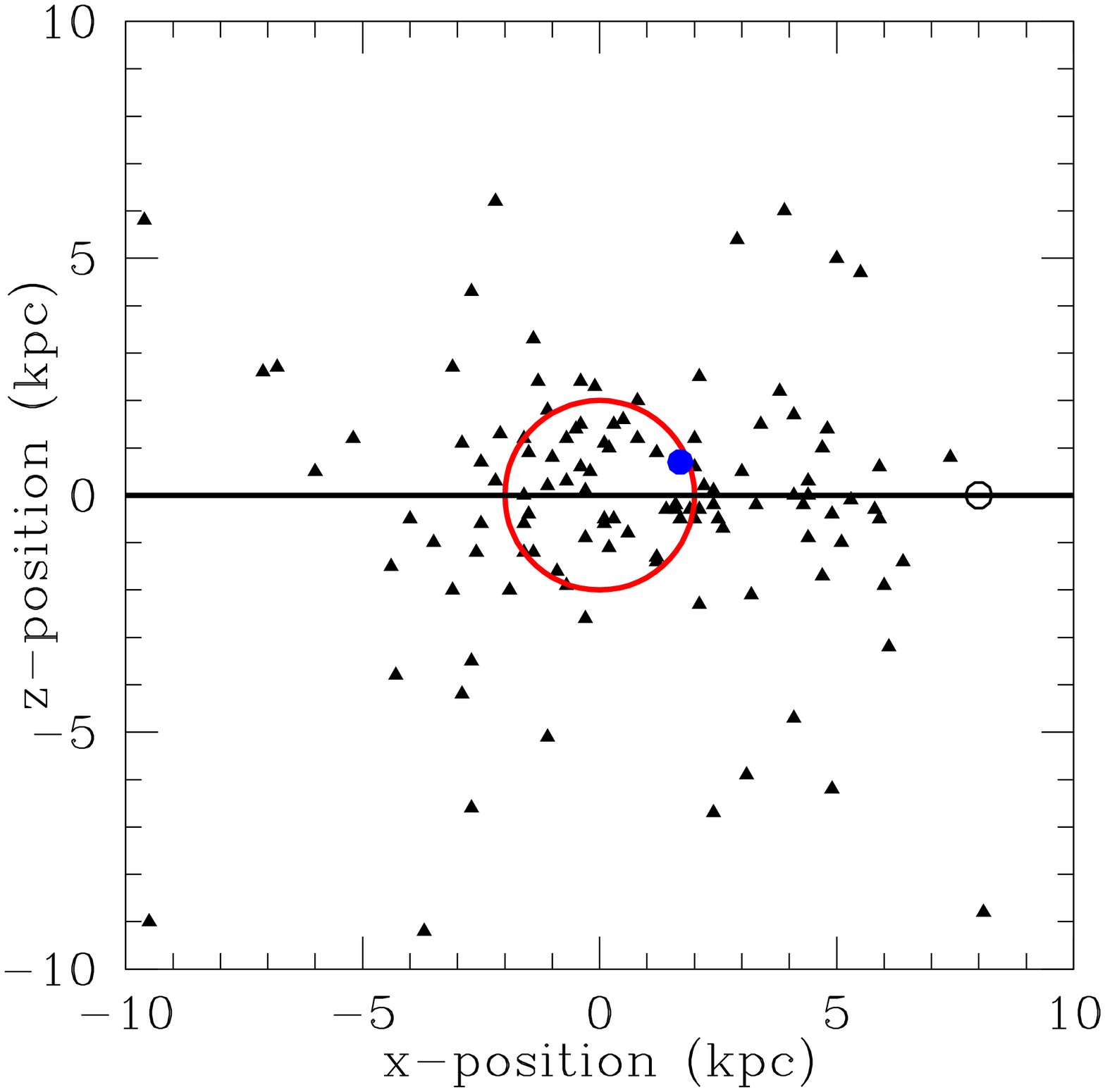}
    \caption{Position of NGC 5927 with respect to the Galactic Center. Filled blue circle represent NGC 5927 and filled black triangles GCs from \citet[2010 ed.]{Harris1996}. Dashed line is the distance from the Galactic Center to the Sun R$_{\sun}$=8 kpc, red circle is the Bulge radius with R$_{bulge}$=2 kpc and the black line is the Galactic plane. \textit{Upper panel}: Represent X-Y plane and \textit{Lower panel}: X-Z plane.}
     \label{Fig:pos}
\end{figure}

\begin{table*}
\caption{Chemical abundances derived from our FLAMES/UVES data. First column: Elements ID. Columns 2-8: Abundances derived from our observed stars. Column 9: NGC 5927 mean abundance. Column 10: Solar abundances adopted for the analysis.}
 \begin{large}
\begin{tabular}{lccccccccc}
\hline
\hline
El&		1&		2&	3&	4&	5&	6&	7&	Cluster$^1$&	 Sun\\
\hline
$[O/Fe]$&			0.4& 0.08& 0.39& 0.24& 0.13& 0.18& 0.18&			0.23$\pm$	0.05&	8.83\\
$[Na/Fe]_{NLTE}$&	0.07& 0.33& 0.24& 0.24& 0.29& 0.08& -0.02&		0.18$\pm$	0.05&	6.32\\
$[Mg/Fe]$&			0.30& 0.30& 0.22& 0.25& 0.29& 0.22& 0.30&			0.27$\pm$	0.02&	7.55\\
$[Al/Fe]$&			0.48& 0.55& 0.35& 0.59& 0.56& 0.44& 0.45&			0.49$\pm$	0.03&	6.43\\
$[Si/Fe]$&			0.24& 0.13& 0.29& 0.12& 0.31& 0.24& 0.32&			0.24$\pm$	0.03&	7.61\\
$[Ca/Fe]^*$&		0.19& 0.15& 0.08& 0.15& 0.15& 0.17& 0.13&			0.15$\pm$	0.01&	6.39\\
$[Sc/Fe]$&			0.35& 0.32& 0.23& 0.34& 0.32& 0.41& 0.29&			0.32$\pm$	0.02&	3.12\\
$[Ti/Fe]$&			0.36& 0.3& 0.26& 0.33& 0.31& 0.41& 0.24&			0.32$\pm$	0.02&	4.94\\
$[V/Fe]^*$&			0.45& 0.37& 0.23& 0.41& 0.36& 0.45& 0.3&			0.37$\pm$	0.03&	4.00\\
$[Cr/Fe]^*$&		0.06& -0.01& -0.02& 0.08& 0.05& 0.04& 0.01&		0.03$\pm$	0.02&	5.63\\
$[Mn/Fe]$&			-0.14& 0.04& -0.04& -0.05& -0.11& -0.14& -0.22&	-0.09$\pm$	0.03&	5.37\\
$[Fe/H]^*$&			-0.54& -0.49& -0.44& -0.47& -0.51& -0.37& -0.49&	-0.47$\pm$	0.02&	7.50\\
$[Co/Fe]^*$&		0.35& 0.36& 0.3& 0.42& 0.39& 0.49& 0.33& 			0.38$\pm$	0.03&	4.93\\
$[Ni/Fe]^*$&		0.12& 0.18& 0.13& 0.14& 0.18& 0.25& 0.22& 		0.17$\pm$	0.02&	6.26\\
$[Cu/Fe]$&			0.3& 0.32& 0.08& 0.44& 0.33& 0.43& 0.31&			0.32$\pm$	0.05&	4.19\\
$[Zn/Fe]$&			-0.06& -0.03& 0.01& -0.21& -0.14& 0.11& 0.03&		-0.04$\pm$	0.04&	4.61\\
$[Y/Fe]$&			-0.06& -0.09& -0.28& -0.16& -0.11& -0.03& -0.31&	-0.15$\pm$	0.04&	2.25\\
$[Zr/Fe]$&			0.14& -0.08& 0.2& 0.03& 0.02& -0.06& -0.08&		0.02$\pm$	0.05&	2.56\\
$[Ba/Fe]$&			0& -0.06& -0.13& -0.17& -0.03& -0.09& -0.08&		-0.08$\pm$	0.02&	2.34\\
$[Ce/Fe]$&			-0.26& -0.27& -0.2& -0.11& -0.21& -0.28& -0.3&	-0.23$\pm$	0.03&	1.53\\
$[Nd/Fe]$&			0.36& 0.16& 0.16& 0.21& 0.16& 0.13& 0.16&			0.19$\pm$	0.03&	1.59\\
$[Eu/Fe]$&			0.55& 0.29& 0.56& 9.99& 9.99& 0.36& 0.38&			0.43$\pm$	0.04&	0.52\\
\hline
\end{tabular}
 \end{large}
\begin{tabular}{lccccccccc}
$^1$The errors are statistical errors obtained from the mean.& & & & & & & & &\\
$^*$Elements measured by EWs.& & & & & & & & &\\
\end{tabular}
\label{tab:abundance}
\end{table*}

\begin{table*}
\caption{Columns 2-5: Differences on abundance measurements due to the variations of each atmospheric parameter. Column 6: Differences on abundance measurements due to the S/N. Column 7: Estimated total error due to atmospheric errors. Column 8: Observed errors.}
 \begin{large}
\begin{tabular}{lccccccc}
\hline
\hline
ID&	$\Delta T_{eff}=65K$& $\Delta log(g)=0.09$& $\Delta v_t=0.08$& $\Delta [Fe/H]=0.03$& S/N& $\sigma_{tot}$& $\sigma_{obs}$ \\
\hline
$\Delta ([O/Fe])$&		0.01&	0.03&	0.00&	0.03&	0.02&	0.05&	0.12\\
$\Delta ([Na/Fe])$&	0.02&	-0.05&	-0.05&	-0.02&	0.02&	0.08&	0.13\\
$\Delta ([Mg/Fe])$&	0.03&	0.00&	-0.05&	0.01&	0.03&	0.07&	0.04\\
$\Delta ([Al/Fe])$&	0.01&	-0.02&	-0.03&	-0.01&	0.03&	0.05&	0.08\\
$\Delta ([Si/Fe])$&	-0.02&	0.00&	-0.03&	0.03&	0.02&	0.05&	0.08\\
$\Delta ([Ca/Fe])$&	0.08&	-0.01&	-0.02&	0.01&	0.06&	0.10&	0.03\\
$\Delta ([Sc/Fe])$&	0.01&	0.01&	0.01&	0.01&	0.02&	0.03&	0.06\\
$\Delta ([Ti/Fe])$&	0.10&	0.00&	-0.03&	0.00&	0.05&	0.12&	0.06\\
$\Delta ([V/Fe])$&		0.11&	-0.01&	-0.04&	-0.01&	0.04&	0.12&	0.08\\
$\Delta ([Cr/Fe])$&	0.06&	0.00&	-0.01&	0.00&	0.04&	0.07&	0.04\\
$\Delta ([Mn/Fe])$&	0.08&	-0.02&	-0.02&	0.00&	0.03&	0.09&	0.08\\
$\Delta ([Fe/H])$&		0.04&	0.01&	-0.03&	0.02&	0.02&	0.06&	0.05\\
$\Delta ([Co/Fe])$&	0.04&	0.02&	-0.03&	0.02&	0.08&	0.10&	0.06\\
$\Delta ([Ni/Fe])$&	0.03&	0.02&	-0.04&	0.03&	0.07&	0.09&	0.05\\
$\Delta ([Cu/Fe])$&	-0.02&	-0.02&	-0.02&	-0.02&	0.05&	0.06&	0.12\\
$\Delta ([Zn/Fe])$&	-0.01&	0.03&	-0.04&	0.04&	0.05&	0.08&	0.11\\
$\Delta ([Y/Fe])$&		-0.04&	0.01&	-0.05&	0.02&	0.05&	0.08&	0.11\\
$\Delta ([Zr/Fe])$&	0.03&	-0.01&	-0.09&	0.02&	0.02&	0.10&	0.11\\
$\Delta ([Ba/Fe])$&	0.05&	0.00&	-0.02&	0.04&	0.03&	0.07&	0.06\\
$\Delta ([Ce/Fe])$&	-0.02&	0.01&	0.01&	0.03&	0.05&	0.06&	0.07\\
$\Delta ([Nd/Fe])$&	0.02&	0.00&	-0.01&	0.00&	0.04&	0.05&	0.08\\
$\Delta ([Eu/Fe])$&	0.00&	0.02&	0.00&	0.03&	0.03&	0.05&	0.08\\
\hline
\end{tabular}
 \end{large}
\label{tab:errors}
\end{table*}

\section{ABUNDANCE ANALYSIS}


Because NGC 5927 is close to the Galactic plane as show in Figure \ref{Fig:pos} ($b=4.86^{\circ}$), its extinction is relatively significant. \citet{Heitsch1999} found a total reddening of $E_{V-I}=0.43\pm0.02$ and a maximum differential reddening of $\Delta E^{MAX}_{V-I}=0.27$. In order to avoid errors introduced by photometric estimations due to the strong extinction, we derive stellar parameters using spectra.

We used the Local Thermodynamic Equilibrium (LTE) MOOG program \citep{Sneden1973} for the abundance analysis coupled with ATLAS9 \citep{Kurucz1970} atmospheric models. For the analysis we used spectral linelists from \citet{Villanova2011} and the solar abundances listed in column 10 of Table \ref{tab:abundance}. 

\begin{figure}
  \centering
    \includegraphics[width=1\linewidth]{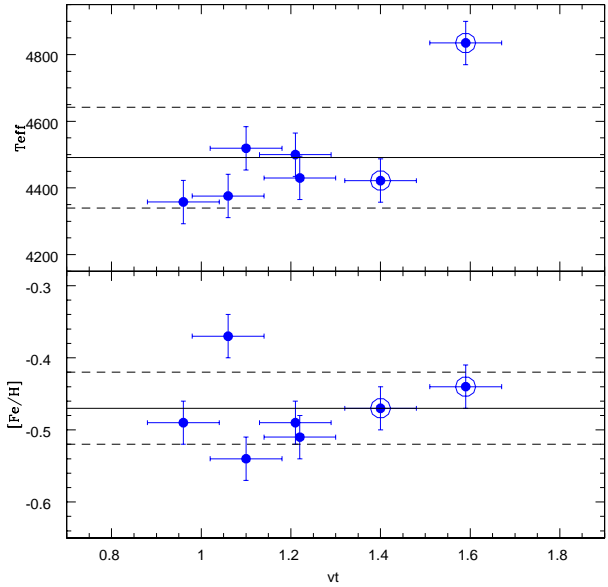}
    \caption{Stellar parameters of the star sample of NGC 5927 in blue, Continuous line indicate the mean value and the dashed line the standard deviation. \textit{Upper panel}: \teff vs \vt. \textit{Lower panel}: [Fe/H] vs \vt.}
     \label{Fig:stellars_params}
\end{figure}

The atmospheric parameters listed in Table \ref{Fig:stellars_params}, were obtained varying the effective temperature ($T_{eff}$), micro-turbulent velocity ($v_t$) and gravity ($log(g)$) in order to remove any trend between the excitation potential and equivalent width vs. abundance, and to satisfy the ionization equilibrium respectively. The FeI and FeII were used for this latter purpose. The \iron value of the model was changed at each iteration according to the output of the abundance analysis. Stellar parameters for each star used in the analysis are listed in Table \ref{tab:stellar_params}. In addition, figure \ref{Fig:stellars_params} shows the stellar parameters T$_{eff}$ and [Fe/H] vs $v_t$. From the upper panel of the figure, star \#3 has the biggest v$_t$ value due to it's high temperature, the hottest star in our sample. In the lower panel we see no trend between [Fe/H] and v$_t$.

\begin{figure}
  \centering
    \includegraphics[width=1\linewidth]{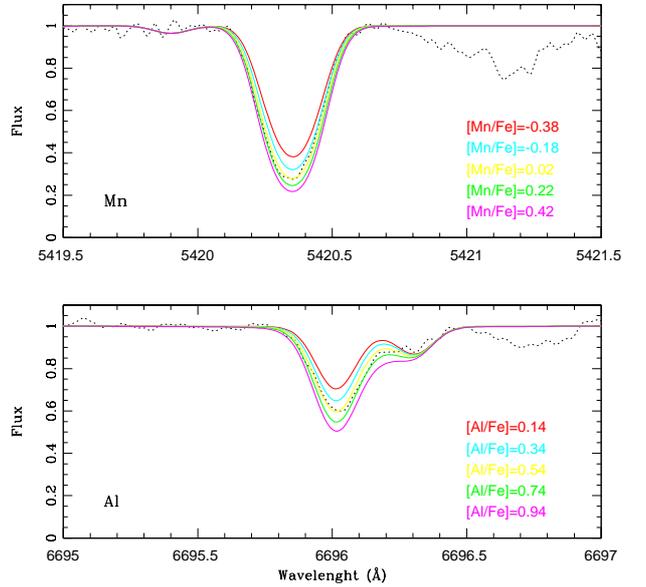}
    \caption{Manganese and Aluminum lines fitting process in Star $\#2$. Dashed lines correspond to observed lines and continuous color lines are synthesised spectra for different abundances.}
     \label{Fig:synth_example}
\end{figure}

Two different techniques were used to obtain the element abundances from our spectra. The Ca, Ti V, Cr, Fe, Co and Ni abundances were measured using Equivalent Widths (\citep[see EWs]{Marino2008}). For O, Na, Mg, Al, Si, Sc, Mn, Cu, Zn, Y, Zr, Ba, Ce, Nd and Eu we used the spectrum-synthesis method because of the blending with other spectral lines. For each line we calculated 5 synthetic spectra varying the abundance, the value that minimizes the rms to fit the observations is our final estimation. Figure \ref{Fig:synth_example} is an example of the fitting process in Star $\#2$ for Mn and Al absorption lines. We selected absorption lines not contaminated by telluric lines and we made special considerations for Ba and Na. We took hyperfine splitting into account only for the Ba measurements since this line is strong enough to consider it. For Na abundances were corrected for non-LTE (NLTE) using corrections provided by the INSPEC \footnote{\url{http://inspect.coolstars19.com/index.php?n=Main.HomePage}} database.

In addition, a detailed internal error analysis was performed by varying \teff , \logg, [Fe/H], and \vt \space by an amount equal to the estimated internal error in each parameter. Then abundances in star $\#2$ was re-measured, assuming that it represents the entire sample. The parameters were varied by $\Delta T_{eff}=+65$ K, $v_t=+0.08$ km s$^{-1}$, $\Delta \log (g)=+0.09$ and $\Delta[Fe/H]=+0.03$ dex. This estimation of the internal errors for atmospheric parameters was performed as in \citet{Marino2008}. The results of errors are listed in Table \ref{tab:errors}, including errors due to the noise in the spectra. This error was obtained for elements whose abundance was obtained by EWs, as the average value of the errors of the mean given by MOOG, and for elements whose abundance was obtained by spectrum-synthesis, as the error given by the fitting procedure. The $\sigma_{tot}$ is the squared sum of the single errors, while $\sigma_{obs}$ is the mean observed dispersion.

\section{RESULTS}

This is the first chemical abundance study of NGC 5927 using high resolution spectra that covers a wide sample of elements (see Table \ref{tab:abundance}) which are involved in different nucleosynthetic processes. The main purpose is to study chemically the evolution of the cluster and investigate possible MPs. For the discussion, field stars from the Bulge, Halo and Thick Disk are compared with our results. We are looking for possible features that could suggest a link between the cluster with any of these structures. Also, we compare our sample set with Bulge GC that have similar metallicities in order to compare their chemical behaviour with NGC 5927.

In the following sections, we present chemical abundances for iron-peak elements, $\alpha$ elements, light elements and heavy elements.

\subsection{Iron-peak elements} \label{sect:iron-peak}

The mean iron abundance value that we found is:

$$[Fe/H]=-0.47 \pm0.02$$

 Reported error is the error on the mean. The spread of iron values are within the errors (see Table \ref{tab:errors}) indicating no intrinsic spread on iron abundances.
 
A slightly different iron content of [Fe/H]=-0.39 dex was found by \citetalias{Pancino2017} using UVES and GIRAFFE data. This difference in iron estimation between \citetalias{Pancino2017} and us is probably due to (1) different linelists used for the abundance analysis, (2) the different processing treatment and (3) the lower resolution of GIRAFFE spectra.

\begin{figure}
  \centering
    \includegraphics[width=0.95\linewidth]{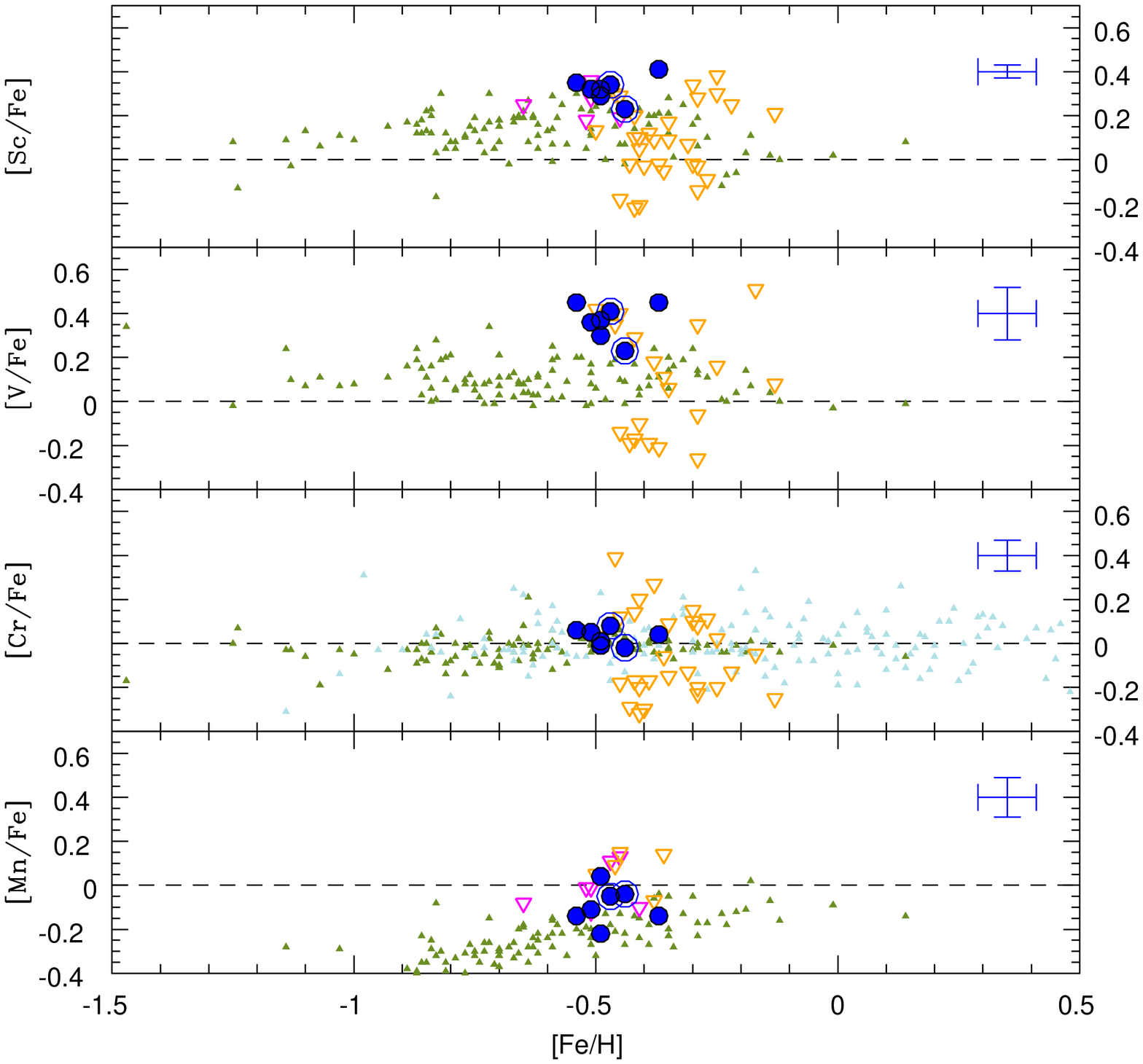}	
    \includegraphics[width=0.95\linewidth]{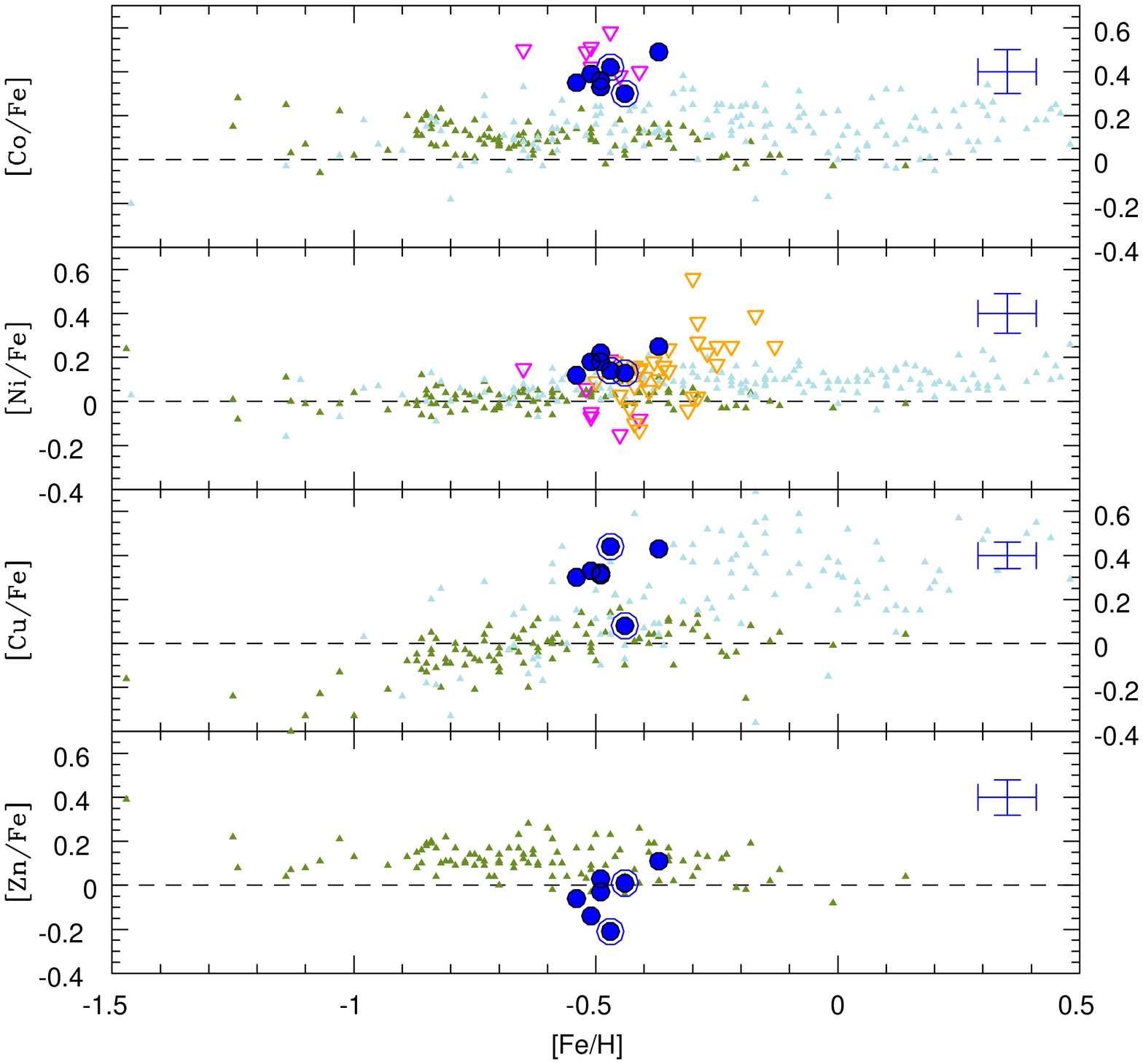}
    \caption{The Sc, V, Cr, Mn, Co, Ni, Cu, and Zn in NGC 5927 represented as filled Blue points shows NGC 5927 stars.  Filled green triangles: Thick Disk stars \citep{Reddy2006}, light blue: bulge field stars \citep{Johnson2014}. Open triangles are bulge GCs, orange: NGC 6441 \citep{Gratton2006,Gratton2007} and magenta: NGC 6440 \citep{Muñoz2013}.}
     \label{Fig:iron}
\end{figure}

 We considered Sc, V, Cr, Mn, Co, Ni, Cu, and Zn as iron-peak elements and their values are listed in Table \ref{tab:abundance}. The Figure \ref{Fig:iron} show each iron-peak element compared with field stars from Thick Disk, Bulge field stars and Bulge GCs found in literature. Some of these elements (Sc, V, Co, Ni, Cu) show an interesting behaviour which relate the cluster to the Bulge rather than the Thick Disk.  This opens the discussion about the origins of NGC 5927, which is linked kinematically to the Thick Disk but under the chemical perspective, shows similarities with the Bulge structure and follows the same behaviour as a Bulge GC.

No large spreads are observed in the iron-peak elements excepting star $\#3$ in Cu. This star is considerably under-abundant having a value of [Cu/Fe]=0.08 dex comparable to the sun, far away from the observe range trend in Cu seen in its companions.
\FloatBarrier

\subsection{$\alpha$ elements} \label{sect:alpha}

The $\alpha$ enhancement was obtained as the mean of Mg, Si, Ca and Ti. The O abundances are considered later as part of the Na-O anti-correlation. The mean value found for $\alpha$ elements is:

$$[\alpha/Fe]=0.25\pm 0.08$$

We also found a mean abundance for pure $\alpha$ elements of $[Ca+Si/Fe]=0.19$ which is lower but consistent for the main value.

\begin{figure}
  \centering
    \includegraphics[width=0.95\linewidth]{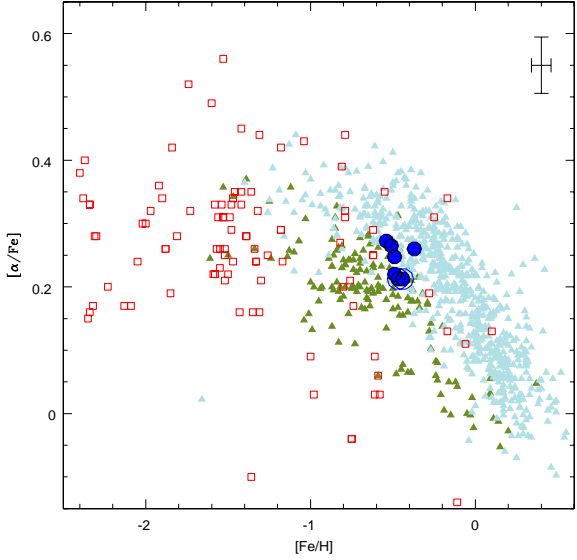}
    \caption{NGC 5927 is represented in filled blue circles. Filled triangles are different samples. Green: Thick Disk stars \citep{Reddy2006}, light blue: Bulge field stars from \citet{Gonzalez2011} and open red squares: GGCs \citep{Pritzl2005}.}
    \label{fig:alpha}
\end{figure}

All the $\alpha$ elements in GGCs are over-abundant compared to the sun as can be seen in Figure \ref{fig:alpha} and NGC 5927 shares this behaviour too as expected. This $\alpha$ element over-abundance is mainly the product of SNeII when massive stars exploded. Type II Supernovae are effective alpha-elements polluters who enrich the intra-cluster gas\citep{Tensley1979,Sneden2004}.

The difference in $\alpha$ elements enhancement between the Thick Disk and Bulge are still under discussion and seems to be that both structures share the same $\alpha$-enhancements. \citet{Bensby2010} compared type F and G dwarf field stars from the Disk and Bulge arguing that they are chemically similar. We did not include data from \citet{Bensby2010} for comparison due to the different type of star used for the study. Our comparison takes into account only giants because in our sample we have this kind of s3tars. In addition, \citet[Figure 16]{Rojas-Arriagada2017} measured Mg for a considerable sample of star from the Disk and Bulge, finding no large difference between them.

\begin{figure*}
  \centering
    \includegraphics[width=0.8\linewidth]{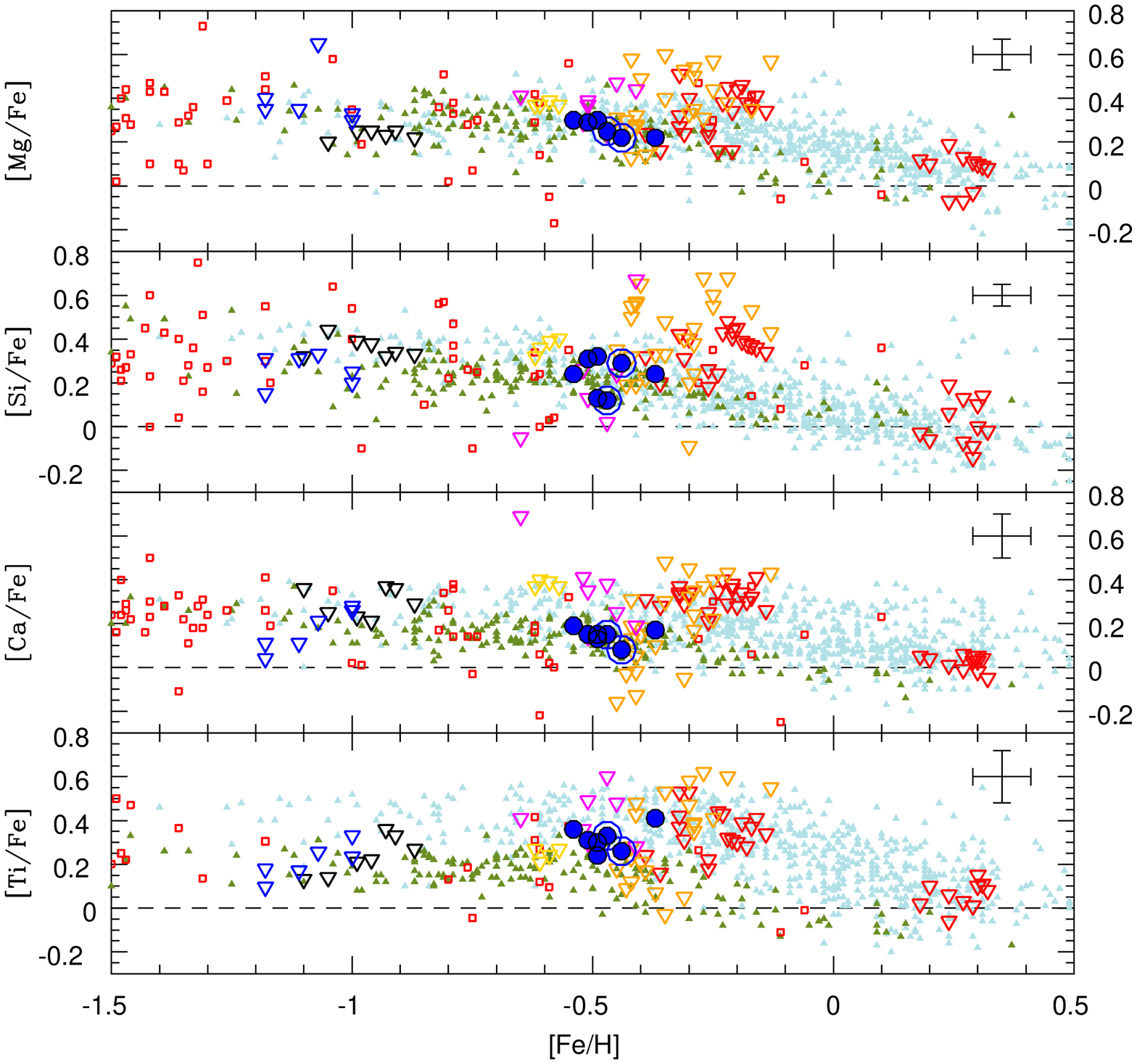}
    \caption{[Mg/Fe], [Si/Fe], [Ca/Fe] and [Ti/Fe] abundances versus [Fe/H] content in NGC 5927, are plotted in filled blue circles. Filled triangles are different samples. Green: Thick Disk stars \citep{Reddy2006} and light blue: Bulge field stars from \citet{Gonzalez2011}.  Red open squares: GGCs \citep{Pritzl2005}. Open triangles represent different Bulge GCs, black: NGC 6723 \citep{Rojas-Arriagada2016}, blue: HP1 \citep{Barbuy2016}, yellow: NGC 6342 \citep{Origlia2005}, orange: NGC 6441 \citep{Gratton2006}, red: Terzan 5 \citep{Origlia2011} and magenta NGC 6440 \citep{Muñoz2013}.}
    \label{fig:alpha_full}
\end{figure*}

In our field stars comparison sample, the $\alpha$ abundances in Bulge shows enhanced values with respect to Thick Disk, but at metallicities of [Fe/H]$\sim$-0.7 they overlap as shown in Figure \ref{fig:alpha}. The same behaviour is seen in each $\alpha$ element Mg, Si, Ca and Ti in Figure \ref{fig:alpha_full}. We can observe in both Figures (\ref{fig:alpha} and \ref{fig:alpha_full}) that NGC 5927 is located just in the middle of the alpha trend of the Thick Disk and Bulge. Also the cluster is compared with other Bulge GCs for each $\alpha$ element.

Besides the position in the Milky Way shown in Figure \ref{Fig:pos}, NGC 5927 has kinematics consistent with being a member of the Thick Disk of the Galaxy \citep{Allen2008,Casetti-Dinescu2007} suggesting it as one of the best Thick Disk candidate GCs. From the chemical point of view, NGC 5927 has $\alpha$-abundances that stand between Bulge and Disk values.

\subsection{Light elements}

In GCs, light elements show variations in their abundances. The responsible processes for these variations are different proton capture chains in previous generation of stars. In the CNO cycle, N is enhanced while C and O are depleted, also in the Ne-Na cycle Ne is depleted and Na enhanced and for the Mg-Al cycle Mg is depleted and Al enhanced. The observation of these abundance differences tell us about the MPs phenomenon. In NGC 5927 we observe a spread in O, Na and Al larger than internal errors indicating a real spread in these elements, and thus confirming that this cluster contains MPs. However, the observed dispersion in Mg is within the range expected from errors alone, suggesting no intrinsic variation (See Table \ref{tab:errors}). In the following, we discuss about the relations between these elements, looking for features in NGC 5927 in order to understand its MPs.

\subsubsection{O-Na anti-correlation}\label{sect:O-Na}

In almost every GC an anti-correlation exist between O and Na and it has been well studied in many of them (e.g. \cite{Carretta2009b}). This anti-correlation is one of the most important chemical signature of MPs since we can differentiate populations along the extension of the anti-correlation (\citet{Carretta2009a}). In the case of NGC 5927, the first study of the O-Na anti-correlation was performed by \citetalias{Pancino2017} and their results are in good agreement with us. The main difference in the Na values in \citetalias{Pancino2017} with respect to our measurements are due to the fact that we take NLTE correction on Na into account. The offset in Sodium between the two datasets is $\sim 0.2$ dex.

\begin{figure}
  \centering
    \includegraphics[width=0.95\linewidth]{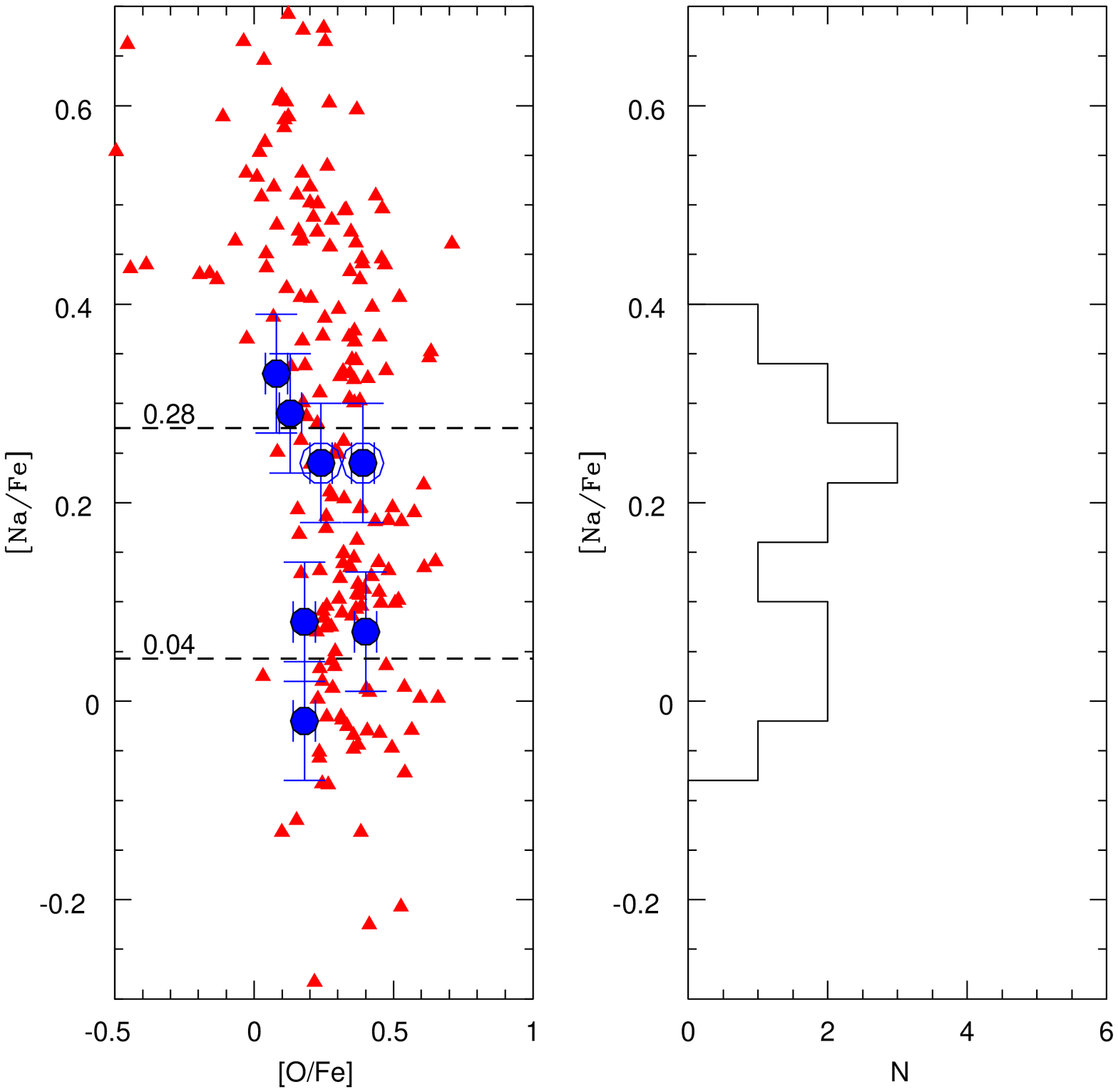}
    \caption{Left: O-Na anti-correlation. Filled blue circles for NGC 5927 and filled red triangles represent 214 red giants from the 19 clusters with UVES  \citep{Carretta2009b}. Right: [Na/Fe] shows a bimodal distribution. The SG observed in the diagram correspond to the $\sim70\%$ which is in good agreement with the generation percentage acording to \citet{Carretta2009b}.}
    \label{fig:O_Na_histo}
\end{figure}

 The Figure \ref{fig:O_Na_histo} shows Na as a function of O next to the Na abundances distribution. In our sample, Na seems to be distributed in two groups suggesting a possible bimodal distribution. Of course, our small sample may not be statistical significant and more data are required to verify this behaviour. The mean [Na/Fe] in Na-rich ($\#2$, $\#3$, $\#4$, $\#5$) and Na-poor ($\#1$, $\#6$, $\#7$) groups is 0.28 dex and 0.04 dex respectively, both with a standard deviation of 0.04. However, this bi-modality can be observed in \citetalias{Pancino2017} (Figure 2) as well.

\begin{figure}
  \centering
    \includegraphics[width=0.95\linewidth]{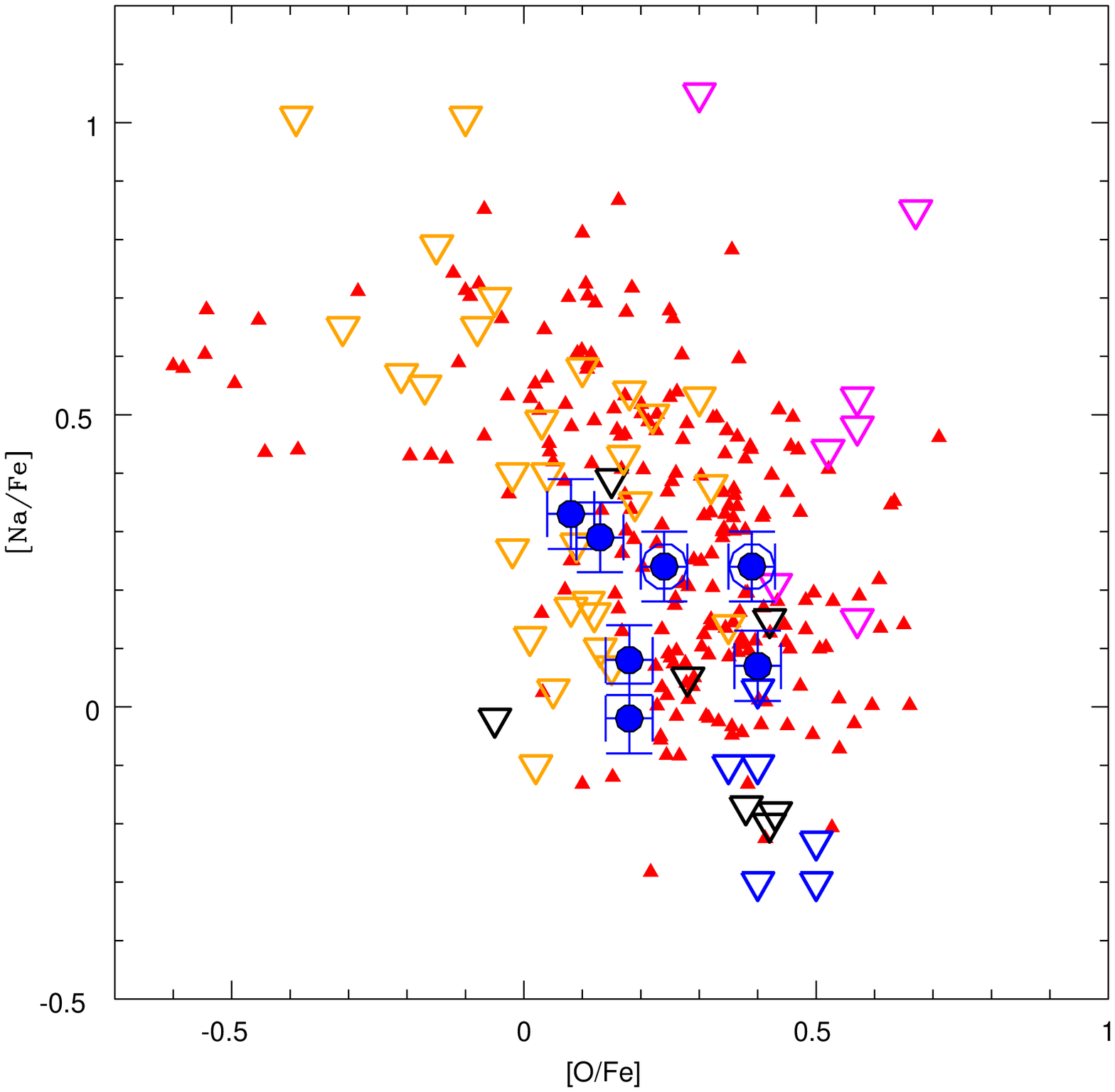}
    \caption{O-Na anti-correlation for NGC 5927 in filled blue circles and filled red triangles represent 214 red giants from 19 clusters with UVES \citep{Carretta2009b}. Open triangles represent different Bulge GCs. Black: NGC 6723 \citep{Rojas-Arriagada2016}, blue: HP1 \citep{Barbuy2016}, orange: NGC 6441 \citep{Gratton2006,Gratton2007} and magenta NGC 6440 \citep{Muñoz2013}.}
    \label{fig:O_Na}
\end{figure}

We also compared our results with GCs stars from \cite{Carretta2009a} and others Bulge GCs with similar metalicities as shown in Figure \ref{fig:O_Na}. We confirm the existence of the O-Na anti-correlation, but not very extended.

\subsubsection{Mg-Al anti-correlation}\label{sec:MgAl}

\begin{figure}
  \centering
    \includegraphics[width=0.95\linewidth]{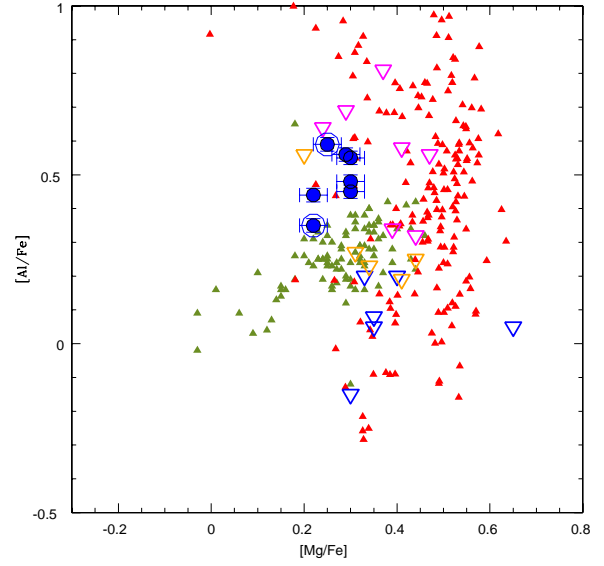}
    \caption{Mg-Al on NGC 5927 in filled blue circles, filled red triangles represent 214 red giants from 19 clusters with UVES \citep{Carretta2009b} and Thick Disk stars \citep{Reddy2006} in green. Open triangles are stars from different GCs. Blue: HP1 \citep{Barbuy2016}, orange: NGC 6441 \citep{Gratton2006} and magenta NGC 6440 \citep{Muñoz2013}.}
    \label{fig:Mg_Al}
\end{figure}

Unlike Na-O, the Mg-Al anti-correlation is only present in some GCs \citep{Carretta2009a}. In the case of NGC 5927, no signs of a clear anti-correlation can be distinguish in Figure \ref{fig:Mg_Al}. In this figure our Mg values shows an under-abundance compared with the trend of \citet{Carretta2009b} mainly because the sample comes from Halo GCs.

Even when NGC 6440 an NGC 6441 share similar metallicity with NGC 5927, both clusters shows an over-abundance in Mg for the majority of the stars but NGC 5927 follow the galactic trend. This behaiviour may be due that NGC 6440 and NGC 6441 are bulge GCs and they follow similar patterns in Mg but not the case of NGC 5927. The values in Al instead, are over-abundant compared with field stars from Thick Disk and enhanced by $\sim 0.2$ dex compared to Mg.

The Mg-Al analysis in NGC 5927 performed by \citetalias{Pancino2017} (Figure 3) include data from UVES and GIRAFFE. They also observe no signs of anti-correlation and the abundances in Mg and Al are in good agreement with us.

\subsubsection{Na-Al correlation}

\begin{figure}
  \centering
    \includegraphics[width=0.95\linewidth]{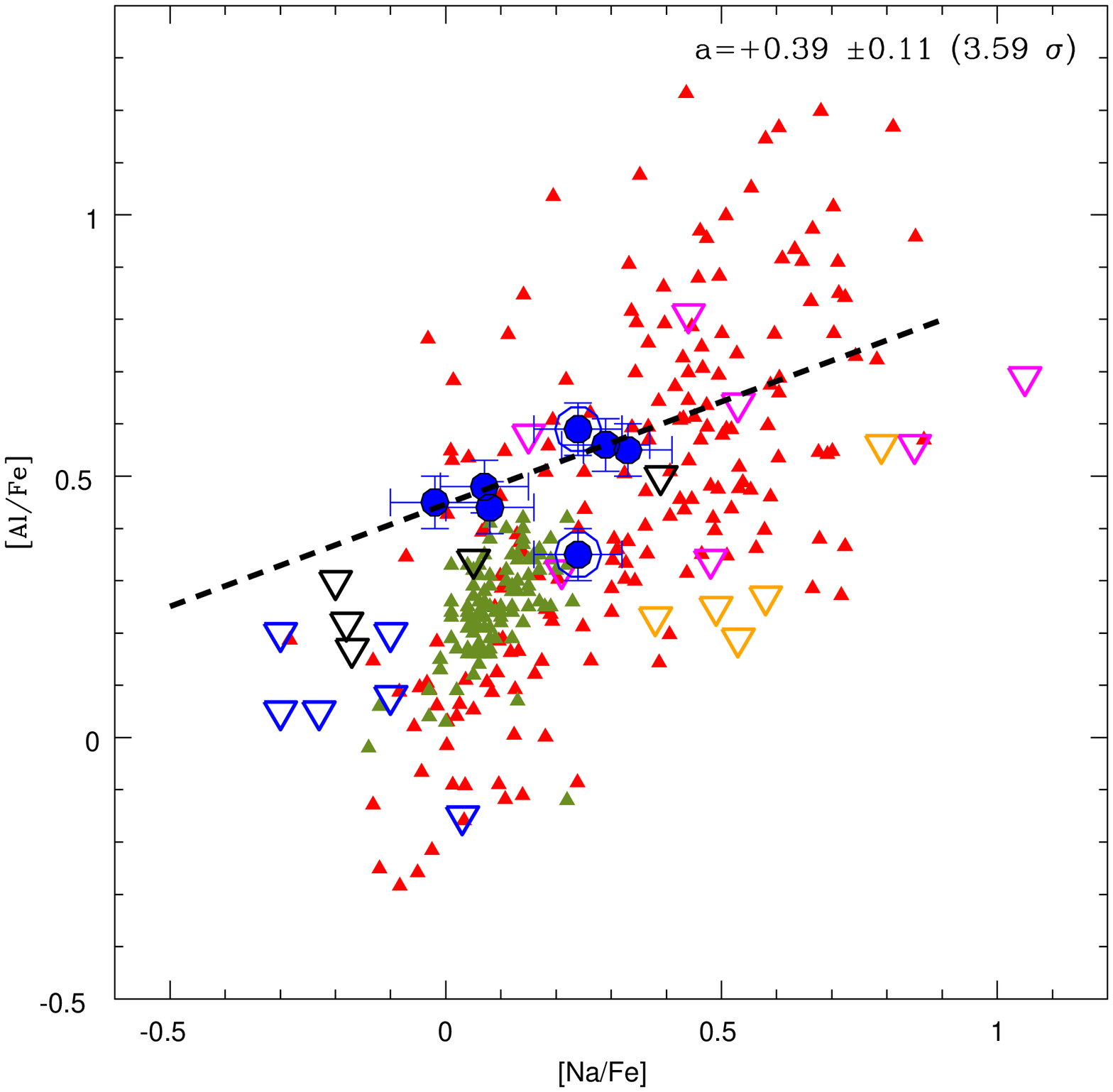}
    \caption{Na-Al correlation for NGC 5927 in filled blue circles and filled red triangles represent 214 red giants from 19 clusters with UVES \citep{Carretta2009b} and Thick Disk stars \citep{Reddy2006} in green. Open triangles represent different Bulge GCs. Black: NGC 6723 \citep{Rojas-Arriagada2016}, blue: HP1 \citep{Barbuy2016}, orange: NGC 6441 \citep{Gratton2006} and magenta NGC 6440 \citep{Muñoz2013}. Dashed is the best fit for NGC 5927 excluding Star $\#3$.}
    \label{fig:Na_Al}
\end{figure}

The correlation between Na and Al in GCs it is probably due to the Ne-Na and Mg-Al chain  activation in previous generations where massive stars enhanced Na and Al respectively. For the case of NGC 5927 we observe a correlation between Na and Al in Figure \ref{fig:Na_Al} even when we do not see a clear anti-correlation in Mg-Al (See Section \ref{sec:MgAl}). The correlation is not continuous due to the Na distribution, grouping it in two groups as is described in Section \ref{sect:O-Na}. The presence of a Na-Al correlation without a Mg-Al anti-correlation has been observed before in other GCs (e.g. M4 by \citet{Marino2008}). We attribute this behaviour in NGC 5927 to a possible activation of the Mg-Al chain reaction in previous generation but with a slow rate.

  The linear fit used for the correlation has an slope of $a=+0.39$ with an error of $\pm 0.11$. This fit was made excluding Star $\#3$ which is the most Al-poor one and shows chemical behaviours (mostly under-abundances) that do not follow the trends described by their companions. The significance of the slope is about 3.5$\sigma$. The anomalies in star $\#3$ as can be seen also in Figure \ref{fig:Mg_Al} for Mg-Al and Figure \ref{Fig:iron} for iron-peak elements (specially in Cu).

\subsection{Heavy elements}

The neutron capture is the nuclear reaction process responsible for the formation of heavy elements. Depending on the neutron capture rate, we can separate  these process in two types: (1) When the capture time is longer than the beta-decay, it is called s-process. (2) If the neutron capture is faster than the beta-decay, we have the r-process \citep{Sneden2008}.

The elements mainly produced by s-process can be light-s elements such as Y, Zr and heavy-s elements like Ba and Ce. This process occurs principally in the AGB phase of intermediate mass stars. On the other hand, the r-process elements are mainly generated during the SNe II explosions. One of the elements produced mainly by this process is Eu which can be used to determine how strong is the contribution of the process in the chemical evolution of the cluster. Also, SNeII produces iron-peak elements and $\alpha$-elements.

\begin{figure}
  \centering
    \includegraphics[width=0.95\linewidth]{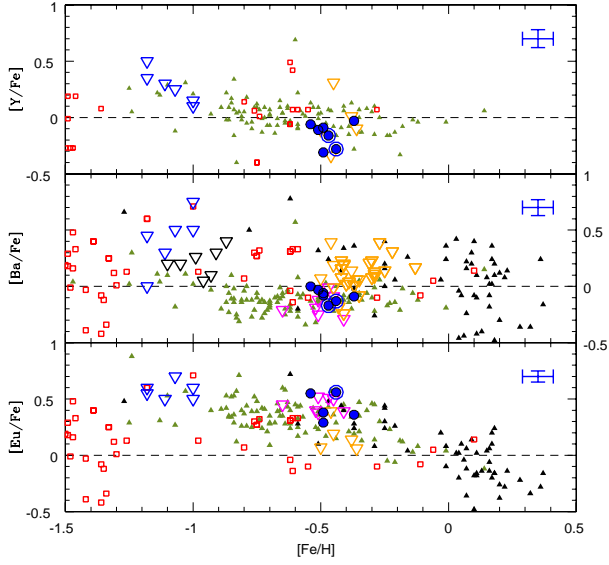}
    \caption{Heavy elements values as function of [Fe/H]. NGC 5927 is represented in filled blue circles. Filled triangles are different samples. Green: Thick Disk stars \citep{Reddy2006}, black: Bulge field stars \citep{Swaelmen2016}. Open red squares are GCs stars \citet{Pritzl2005} Open triangles represent different Bulge GCs, black: NGC 6723 \citep{Rojas-Arriagada2016}, blue: HP1 \citet{Barbuy2016}, orange: NGC 6441 \citep{Gratton2006} and magenta NGC 6440 \citep{Muñoz2013}.}
    \label{fig:heavy}
\end{figure}

In Figure \ref{fig:heavy} we compare Y, Ba and Eu abundances of NGC 5927 with field stars and also with Bulge GCs. For Y and Ba, abundance values are sub-solar and shows some differences with respect to the Bulge GCs, with the exception of NGC 6440. We underline the strong similarities in heavy elements enhancements between NGC 5927 and NGC 6440. These similarities have also been observed in the $\alpha$ and iron-peak elements (Sections \ref{sect:alpha} and \ref{sect:iron-peak} respectively). For Eu instead, values are over-abundant with respect to the Sun. This indicates a large contribution of SNeII to the proto-cluster cloud.

\begin{figure}
  \centering
    \includegraphics[width=0.95\linewidth]{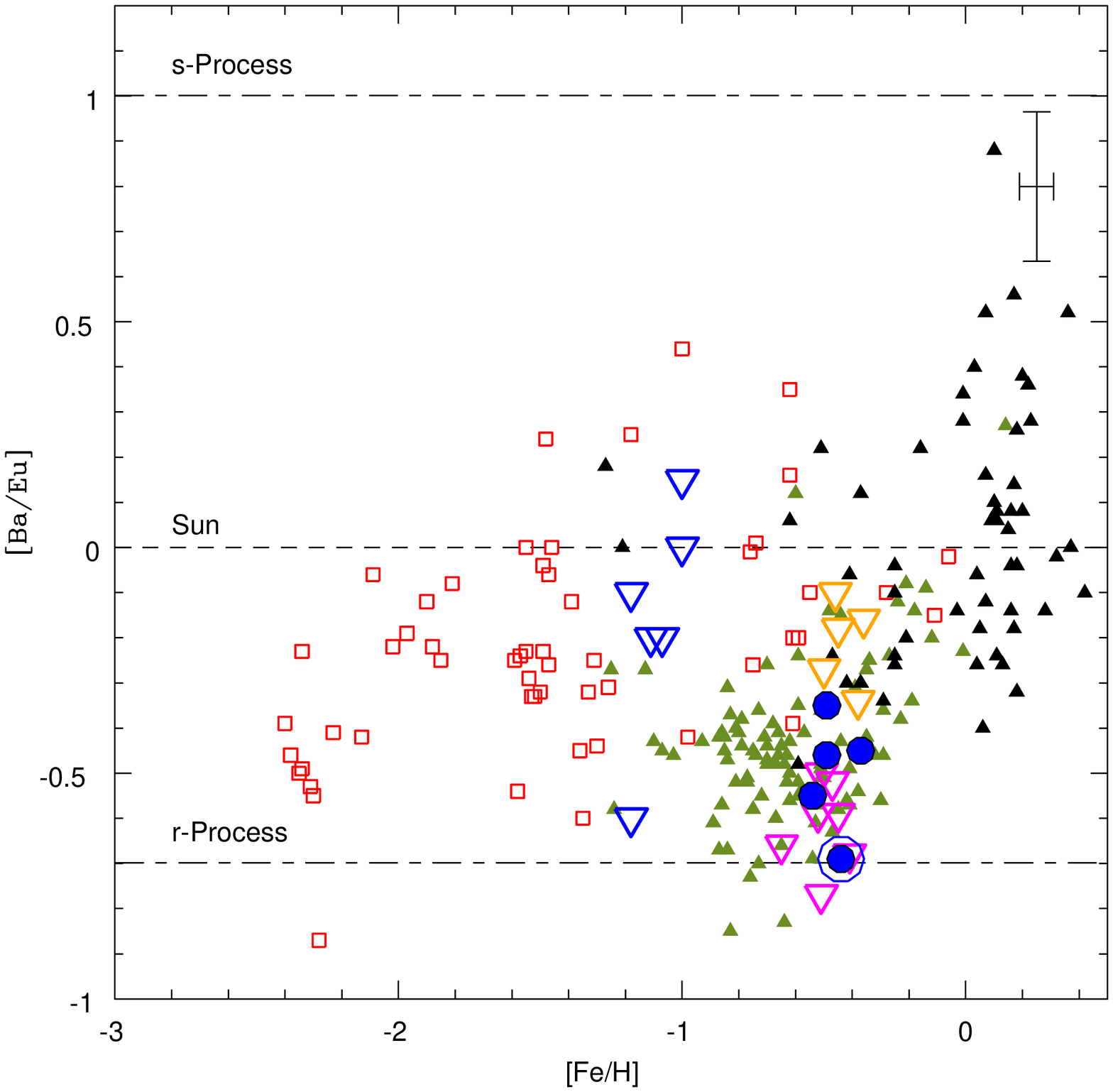}
    \caption{[Be/Eu] ratio as function of [Fe/H]. In filled blue circles, NGC 5927. Filled triangles are different samples. Green: Thick Disk stars \citep{Reddy2006} and black: Bulge field stars \citep{Swaelmen2016}. Open red squares: GGCs \citep{Pritzl2005}. Open triangles represent different Bulge GCs, orange: NGC 6441 \citep{Gratton2006}, blue HP1 \citep{Barbuy2016} and magenta NGC 6440 \citep{Muñoz2013}.}
    \label{fig:BaEu}
\end{figure}

Since Ba is mainly processed by s-process and Eu mainly by r-process, the [Ba/Eu] ratio is an indicator of which process contributed more in the chemical evolution of the cluster. In Figure \ref{fig:BaEu}, we observe that [Ba/Eu] ratio in NGC 5927 is mainly dominated by r-process contributions. This suggest that the early proto-cluster environment was polluted mainly by SNeII. In addition, is notable the similarity (again) between NGC 5927 and NGC 6440 showing values very similar indicating that both cluster share similar evolutionary scenarios.

\section{SUMMARY AND CONCLUSIONS}

In this paper we present a detailed chemical abundances analysis of the Globular Cluster NGC 5927. A total of 22 elements in 7 giant stars are consider for the analysis using high resolution spectroscopy taken by UVES instrument with a moderate S/N  ($\sim40$). We measure 16 elements by spectrum-synthesis and 7 elements by EW, including an accurate error analysis. The results are compared with different components of the Milky Way such Thick Disk, Bulge field stars and Bulge GCs.

We obtained the following results:

\begin{itemize}
\item We found a mean metallicity of [Fe/H]=-0.47 dex with a $\sigma_{obs}=0.05$. This value is in good agreement with the value estimated by \citet[2010 ed.]{Harris1996} ([Fe/H]=-0.49). Also we rule out an intrinsic spread in the iron content.
No significant spread is visible in other iron-peak elements. For some elements, NGC 5927 seem to follow the Bulge trend rather than the Thick Disk.
Previous literature refers to NGC 5927 as one of the most metal-rich Thick Disk GCs but its metallicity is comparable with the Bulge field stars and Bulge GCs (ie. NGC 6440). There is a difference of $\sim0.1$ dex between our measurements and \citet{Pancino2017} ([Fe/H]=-0.39 dex).

\item We confirm the existence of an anti-correlation between O and Na but the spread in O is small. The Na distribution seem to be bi-modal but we caution that our sample size is small. A large sample is required to confirm this behaviour.

\item We observe no clear evidence for Mg-Al anti-correlation in NGC 5927, however we observe a Na-Al correlation. The spread in Al may indicate a low Mg-Al cycle activity. Also, our abundances are in good agreement with \citet{Pancino2017} in this topic.

\item About $\alpha$ content, NGC 5927 matches the typical trends of the Galaxy field stars. We found an alpha enrichment of [$\alpha$/Fe]=0.25 $\pm$0.08.
In each $\alpha$ element (Mg, Si,Ca and Ti), the values for NGC 5927 are not so enhanced as those for Bulge GCs but they still are over-abundant. These enhancements indicate that the cluster has experienced rapid chemical evolution which is related to contributions from SNeII. We decide not to relate NGC 5927 with the Bulge or Disk components using alpha abundances, since whether there exists a difference in alpha abundance between these two components is still under debate.

\item A strong contribution of r-process is suggested by analysing [Ba/Eu] ratios pointing out SNeII as principal polluter for the proto-cluster cloud.

\item The heavy s-process elements shows no significant spread, which seems to contradict the theory that AGB stars are polluters for the second generation stars.

\item We highlight the similarity of NGC 5927 with NGC 6440 Bulge GC in $\alpha$, iron-peak and heavy elements, suggesting similar origins between these two GCs.

\item Star $\#3$ shows anomalies abundances with respect to the other stars of the sample. Even when this star is one of the AGB stars in our sample, we attribute this behaviour to some problem with the spectrum of star $\#3$. Also exist a very low probability, but still possible, that this star may be a field star, even when it passed all the membership criteria.
\end{itemize}

In addition, considering that NGC 5927 (1) is a very old GC with an age between 10.7 Gyr \citet{Vandenberg2013} and  12.25 Gyr \citep{Dotter2010}, as old as the Milky Way itself; (2) its calculated orbit using proper motions \citep{Allen2008} take place between 4 and 6 kpc; (3) the iron-peak enhancement in some elements follows the Bulge trend rather than the Thick Disk and; (4) the chemical similarities of NGC 5927 with Bulge GCs and especially with NGC 6440, lead us to consider a formation scenario where NGC 5927 was formed from material in-between the Bulge and Disk, when both structures were still in formation. The gas in this region of transition, with which NGC 5927 was formed, left chemical traces in the cluster with its properties in-between the Bulge and the Disk observed today.

\section{AKNOWLEDGEMENTS}
A.M gratefully acknowledges the support provided by Direcci\'on de Postgrado UdeC and Chilean BASAL Centro de Excelencia en Astrof\'isica y Tecnolog\'ias Afines (CATA) grant PFB-06/2007 for international conference which helped to develop the paper. SV gratefully acknowledges the support provided  by Fondecyt reg. n. 1170518. C.M. is supported by CONICYT (Chile) through Programa Nacional de Becas de Doctorado 2014 (CONICYT-PCHA/Doctorado Nacional/2014-21141057)

\bibliographystyle{mnras}	

\bibliography{Mura} 

\bsp 
\label{lastpage}

\end{document}